\documentclass[aps,preprint]{revtex4-1}
\usepackage{amsmath,graphicx,bbm,mathrsfs,amssymb,pst-all,bm,color}

\begin{document}

\title{Polynomial-time quantum algorithm for solving the hidden subgroup problem}
\author{Hefeng Wang}
\email{wanghf@mail.xjtu.edu.cn}
\affiliation{MOE Key Laboratory for Nonequilibrium Synthesis and Modulation of Condensed Matter, 
Xi'an Jiaotong University, Xi'an, 710049, China \\
School of Physics, Xi'an Jiaotong University and Shaanxi Province
Key Laboratory of Quantum Information and Quantum Optoelectronic
Devices, Xi'an, 710049, China}

\begin{abstract}
The hidden subgroup problem~(HSP) is one of the most important problems in quantum 
computation. Many problems for which quantum algorithm achieves exponential speedup over
its classical counterparts can be reduced to the Abelian HSP. However, there is no 
efficient quantum algorithm for solving the non-Abelian HSP. We find that the HSP 
can be reduced to a nested structured search problem that is solved efficiently by 
using a quantum algorithm via multistep quantum computation. Then we solve the HSP and 
problems that can be reduced to both the Abelian and the non-Abelian HSP in polynomial 
time by using this algorithm.
\end{abstract}

\maketitle

\section{Introduction}
Many computational problems can be classified as the constraint satisfaction problems, 
which consist of a number of variables and each of them has a set of domain values, 
together with a set of constraint functions that are required to be satisfied simultaneously, 
e.g. the propositional satisfiability problem and combinatorial optimization problem. 
All possible sets of the variable-value assignments form the computational basis
states~(CBS) of the problem. The solution to a constraint satisfaction
problem is a set of CBS that satisfy all the constraint functions. Classical
structured search algorithm reduces the search space of the problem by
eliminating certain assignments of a subset of the variables that violate
the constraint functions~\cite{struct}. In the case where the variables in
the constraint functions are independent, classical structured search
algorithms can obtain the solution efficiently. The problem can be solved
through the binary tree searching, where the solution is nested in a series
of search spaces with increasing dimension as $2^{0}\subset 2^{1}\subset
\cdots \subset 2^{n}$ in an $n$-bit binary tree, by checking the bits
sequentially, the search space is reduced in a rate of $1/2$ in each step,
and the solution is obtained in $n$ steps. For instance, the $n$-bit $1$-SAT
problem~\cite{Hogg}, where the bits are examined one by one and each clause
eliminates one value of a bit. In general, however, the constraint functions
consist of variables that are coupled together, they cannot be divided or
may be divided into only a few levels, e.g., the $3$-SAT problem, thus the
search space is still exponentially large, and classical structured search
algorithms cannot solve these problems efficiently.

In some cases, a constraint satisfaction problem in which the variables of
the constraint functions are inseparable can be decomposed by using some
oracles and has the nested structure as in the binary tree searching where
the search space of the problem is reduced in a finite rate. Classical
search algorithms have to check the CBS of the problem one by one using the
oracles to find the solution. While in quantum computing, the CBS of a
problem can be prepared in a superposition state as input to the oracles,
and the function values of the CBS are evaluated simultaneously in quantum
parallelism. By using this property of quantum computing, a quantum
algorithm can efficiently solve some problems that are difficult for
classical search algorithms. In Ref.~\cite{wyx}, we proposed an efficient
quantum algorithm that achieves exponential speedup over classical search
algorithms in solving a search problem with nested structure via multistep
quantum computation based on quantum resonant transition.

Our algorithm solves a problem by finding the ground state of the problem
Hamiltonian that encodes the solution to the problem in multiple steps. We
first construct a sequence of intermediate Hamiltonians by decomposing the
problem using some oracles to form a Hamiltonian evolution path from an
initial Hamiltonian to the problem Hamiltonian. Then starting from the
ground state of the initial Hamiltonian, evolving it through ground states
of the intermediate Hamiltonians sequentially to reach the ground state of
the problem Hamiltonian via quantum resonant transition~(QRT)~\cite{whf0,
whf2}. This algorithm can be run efficiently provided that the energy gap
between the ground and the first excited states of each Hamiltonian and the
overlap between the ground states of any two adjacent Hamiltonians are not
exponentially small. In each step, the ground state of the Hamiltonian of
the step is induced from that of the previous step by using the QRT method
and is protected through quantum entanglement, therefore it can be used
repeatedly to obtain the ground state of the next Hamiltonian without making
copies, thus circumventing the restriction of the no-cloning theorem~\cite%
{noclone1, noclone2} to realize multistep quantum computation. The runtime
of the algorithm is summation of the runtime of each step. The idea of the
algorithm is similar to the domino effect, i.e., assuming a domino can push
down another domino with twice of its size, by building a series of dominos
with increasing size of \{$1$, $2^{1},\ldots ,2^{n}$\}, a small domino with
size $1$ can push down a domino of size $2^{n}$ in $n$ steps.

The hidden subgroup problem~(HSP) is one of the most important problems in
the field of quantum computation. Many important problems can be reduced to
the HSP, e.g., Simon's problem, the factoring problem and the discrete
logarithm problem can be reduced to the Abelian HSP~\cite{nc}, the graph
isomorphism problem~(GIP) and the poly($n$)-unique shortest
vector problem~(SVP) can be reduced to the non-Abelian HSP.
Quantum algorithms can achieve exponential speedup over classical algorithms
for solving the Abelian HSP, however, there is no efficient quantum
algorithm for solving the non-Abelian HSP so far. By analyzing the spectrum
structure of the HSP, we find that the HSP can be decomposed by using a
number of oracles that are constructed based on a sequence of threshold
values, and reduced to the nested structured search problem, therefore it
can be solved efficiently.

\section{The nested structured search problem}
The nested structured search problem~\cite{wyx} is a search problem that contains $N$ items 
with one target item, and can be decomposed by using $m$~[$O(\log N)$] oracles to 
construct $m$ Hamiltonians
\begin{equation}
H_{P_{i}}=-\sum_{\eta _{i}\in A_{i}}|\eta _{i}\rangle \langle \eta _{i}|%
\text{, \ \ }i=1,\ldots ,m
\end{equation}%
and%
\begin{equation}
H_{P_{m}}=H_{m}=H_{P}=-|\eta \rangle \langle \eta |,
\end{equation}%
where the set $A_{i}$ contains $N_{i}$ marked items in the $N$ items and $%
|\eta _{i}\rangle $ are the marked states associated with the marked items.
These sets have a nested structure as $A_{m}\subset A_{m-1}\subset \cdots
\subset A_{1}$, with sizes $N_{m}=1$, $N_{m-1}$, $\ldots $, $N_{1}$,
respectively. The ratio $N_{i}/N_{i-1}$ is finite, where $N_{0}=N$. In
finding the target state $|\eta \rangle $ in the set $A_{m}$, we construct a
Hamiltonian evolution path as%
\begin{equation}
H_{i}=\frac{N_{i}}{N}H_{0}+\left( 1-\frac{N_{i}}{N}\right) H_{P_{i}},\text{
\ \ }i=0,1,\ldots ,m-1,
\end{equation}%
to reach the problem Hamiltonian $H_{P}$, where $H_{0}=-|\psi _{0}\rangle
\langle \psi _{0}|$ and $|\psi _{0}\rangle =\frac{1}{\sqrt{N}}%
\sum_{j=0}^{N-1}|j\rangle $. Both the energy gap between the ground and the
first excited states of each Hamiltonian, and the overlap between the ground
states of any two adjacent Hamiltonians are polynomially large since $%
N_{i}/N_{i-1}$ ($i=1,\ldots ,m$) are finite~\cite{wyx}. The algorithm
becomes simpler if $N_{i}$ are known, then the ground state eigenvalue of $%
H_{i}$ and the overlap $d_{0}^{(i)}=\langle \varphi _{0}^{(i-1)}|\varphi
_{0}^{(i)}\rangle $ can be calculated analytically~\cite{wyx}, where $%
|\varphi _{0}^{\left( i-1\right) }\rangle $ and $|\varphi _{0}^{\left(
i\right) }\rangle $ are the ground states of $H_{i-1}$ and $H_{i}$,
respectively. By using the Hamiltonians $H_{P_{i}}$ sequentially in each
step to construct the Hamiltonian $H_{i}$, the search space of the problem
is narrowed in a finite rate, and the solution state to the problem
Hamiltonian is obtained step by step.

We prepare ($n+1$) qubits including a probe qubit and an $n$-qubit register $%
R$ representing the problem, and describe the algorithm by assuming $N_{i}$
are known since this is the case for solving the HSP. In the $i$th step of
the algorithm, given $H_{i-1}$, its ground state eigenvalue $E_{0}^{\left(
i-1\right) }$ and $|\varphi _{0}^{\left( i-1\right) }\rangle $, and $H_{i}$
and its ground state eigenvalue $E_{0}^{\left( i\right) }$, we are to
prepare the ground state $|\varphi _{0}^{\left( i\right) }\rangle $ of $%
H_{i} $. The algorithm Hamiltonian of the $i$th step is
\begin{equation}
H^{\left( i\right) }=-\frac{1}{2}\omega \sigma _{z}\otimes
I_{N}+H_{R}^{\left( i\right) }+c\sigma _{x}\otimes I_{N},
\end{equation}%
where%
\begin{equation}
H_{R}^{\left( i\right) }=\alpha _{i}|1\rangle \langle 1|\otimes
H_{i-1}+|0\rangle \langle 0|\otimes H_{i}\mathbf{,}\text{\ }i=1,2,\cdots ,m,
\end{equation}%
$I_{N}$ is the $N$-dimensional identity operator, and $\sigma _{x}$ and $%
\sigma _{z}$ are the Pauli matrices. The first term in Eq.~($4$) is the
Hamiltonian of the probe qubit, the second term contains the Hamiltonian of
the register $R$ and describes the interaction between the probe qubit and $%
R $, the third term is a perturbation with $c\ll 1$. The initial state of
the circuit is set as $|1\rangle |\varphi _{0}^{\left( i-1\right) }\rangle $%
, which is an eigenstate of $H_{R}^{\left( i\right) }$ with eigenvalue $%
\alpha _{i}E_{0}^{\left( i-1\right) }$. We set $\alpha _{i}=\left(
E_{0}^{\left( i\right) }-\omega \right) /E_{0}^{\left( i-1\right) }$ such
that $E_{0}^{\left( i\right) }-\alpha _{i}E_{0}^{\left( i-1\right) }=\omega $%
, satisfying the condition for resonant transition between the probe qubit
and the transition between states $|\varphi _{0}^{\left( i-1\right) }\rangle
$ and $|\varphi _{0}^{\left( i\right) }\rangle $. The procedure for
obtaining the state $|\varphi _{0}^{\left( i\right) }\rangle $ is: ($i$)
initialize the probe qubit to its excited state $|1\rangle $ and the
register $R$ in state $|\varphi _{0}^{\left( i-1\right) }\rangle $; ($ii$)
implement the time evolution operator $U(t)=\exp \left( -iH^{\left( i\right)
}t_{i}\right) $, where $t_{i}=\pi /\left( 2cd_{0}^{(i)}\right) $; ($iii$)
read out the state of the probe qubit. As the resonant transition occurs,
the system is approximately in an entangled\ state $\sqrt{1-p_{0}^{\left(
i\right) }}|1\rangle |\varphi _{0}^{\left( i-1\right) }\rangle +\sqrt{%
p_{0}^{\left( i\right) }}|0\rangle |\varphi _{0}^{\left( i\right) }\rangle $
where $p_{0}^{\left( i\right) }=\sin ^{2}\left( ct_{i}d_{0}^{(i)}\right) $
is the decay probability of the probe qubit. By performing a measurement on
the probe qubit, if it decays to its ground state, it means that the system
evolves to state $|0\rangle |\varphi _{0}^{\left( i\right) }\rangle $;
otherwise if the probe stays in its excited state, it indicates the register
$R$ remains in state $|\varphi _{0}^{\left( i-1\right) }\rangle $, then we
repeat\ steps ($ii$)-($iii$)\ until the probe decays. The solution state $%
|\varphi _{0}^{\left( m\right) }\rangle =|\eta \rangle $ is obtained in $m$
steps.

\section{The hidden subgroup problem}
The HSP is~\cite{nc}: let $G$ be a finite group, $X$ be a finite set of integers,
and $f:G\longrightarrow X$ be a function such that there exists a subgroup $K<G$
for which $f$ separates the cosets of $K$, that is, for all $g_{1},g_{2}\in G$, 
$f\left( g_{1}\right) =f\left( g_{2}\right) $ only if $g_{1}K=g_{2}K$.
Given a quantum black box for performing the unitary transform $O_{f}|g\rangle
|h\rangle =|g\rangle |h\oplus f(g)\rangle $, for $g\in G$, $h\in X$, and $%
\oplus $ a binary operation on $X$, using information gained from evaluation
of $f$, find a generating set for the subgroup $K$.

The function $f$ hides the subgroup $K$ and is constant on the cosets of $K$. 
A classical algorithm calls a routine evaluating $f(g)$ once for each
group element $g$, and determines $K$ with $|G|$ calls. Quantum algorithm
explores the periodic structure in the Abelian HSP whose irreducible
representation is one dimensional in group theory, and reduces the cost to $%
O(poly(\log |G|))$ by using the quantum Fourier transform~\cite{childs}. The
non-Abelian HSP is more difficult than the Abelian case, and there is no
efficient quantum algorithm so far, see Ref.~\cite{Jozsa} for details.
\begin{figure}[tbp]
\centering
\includegraphics[width=.5\linewidth]{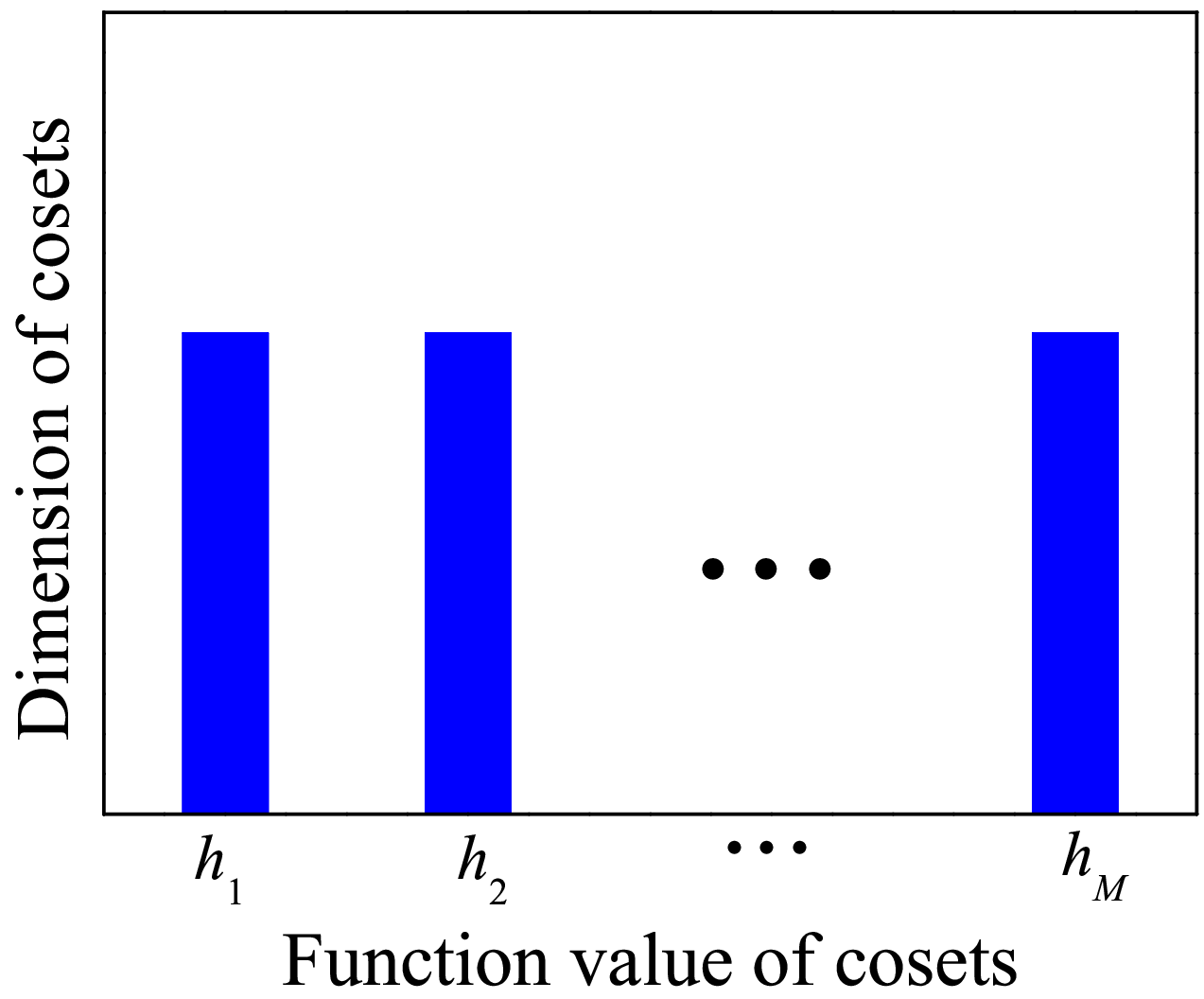}
\caption{Spectrum of the hidden subgroup problem.}
\label{fig:fig1}
\end{figure}

Our algorithm solves the HSP by finding the state of the hidden subgroup
that is mapped to a certain integer, the elements of the subgroup are
obtained by measuring the state. Suppose $f$ maps the subgroup $K=S_{1}$ and
its cosets $S_{2},\cdots ,S_{M}$ to integers $h_{1},h_{2},\cdots ,h_{M}\in X$%
, respectively, where $M=|G|/|K|$, the corresponding coset states are $%
|S_{i}\rangle =\frac{1}{\sqrt{\left\vert K\right\vert }}\sum_{k_{i}\in
S_{i}}|k_{i}\rangle $ on a quantum computer, the goal is to obtain the state
of the hidden subgroup $|K\rangle =|S_{1}\rangle $ that is mapped to $h_{1}$%
. By using the oracle $O_{f}:|j\rangle |0\rangle \longrightarrow |j\rangle
|f\left( j\right) \rangle $, where $|j\rangle $ is the state associated with
the group element $g_{j}\in G$, the Hamiltonian of the HSP is constructed as
\begin{equation}
H_{\text{HSP}}|k\rangle =h_{k}|k\rangle .
\end{equation}%
Without loss of generality, we assume that $h_{k}$ are in increasing order.
Fig.~$1$ shows the spectrum of the HSP, i.e., the dimension of the cosets
v.s. the function values. We can see that the integer $h_{1}$ can be located
in $m=\lceil \log _{2}M\rceil $ steps by using the method of bisection, thus
the corresponding state $|K\rangle $ can be obtained in $m$ steps by using
our algorithm. We prepare a set of threshold values \{$v_{1}=h_{\lfloor
M/2\rfloor }$, $v_{2}=h_{\lfloor M/2^{2}\rfloor },\ldots ,v_{m}=h_{1}$\},
and divide the group elements of $G$ according to their corresponding
eigenvalues by using $m$ oracles based on the threshold values to construct
Hamiltonians $H_{P_{i}}$ as
\begin{equation}
H_{P_{i}}|k\rangle =\Bigg\lbrace%
\begin{array}{c}
\!\!-1\cdot |k\rangle ,\,\mathrm{if}\,h_{k}\leqslant v_{i} \\
\,\,\,\hskip.0003in0\cdot |k\rangle ,\,\,\,\mathrm{if}\,h_{k}>v_{i}\,%
\end{array}%
,\text{ \ \ }i=1,\ldots ,m.
\end{equation}%
This can be achieved by using an oracle that recognizes whether the
eigenvalue $h_{k}$ of a state is larger or less than a threshold value $%
v_{i} $. It is a comparison logic circuitry and can be implemented
efficiently on a quantum computer~\cite[p.264]{nc}~\cite{durr, bari,
grandunif}. The CBS associated with integers that are less than or equal to $%
v_{i}$ form a set $\Pi _{i}$ with size $N_{i}$. They have the nested
structure as $\Pi _{m}\subset \Pi _{m-1}\subset \cdots \subset \Pi _{1}$,
and the ratio $N_{i}/N_{i-1}\approx 1/2$, the set $\Pi _{m}$ contains the
CBS that correspond to the integer $h_{1}$. Then we construct the
intermediate Hamiltonians $H_{i}$ as shown in Eq.~($3$) and run the
algorithm to obtain the state $|K\rangle $ in $m$ steps. The total runtime
of the algorithm for solving the HSP scales as $O(\log |G|)$.

In the following, we apply the algorithm for solving Simon's problem that
belongs to the Abelian HSP, and the GIP and the poly($n$)-unique SVP that
are reduced to the non-Abelian HSP of the symmetric group and the dihedral group,
respectively. The factoring problem, the discrete logarithm problem,
the period-finding problem and the order-finding problem are solved in the same 
way~(see appendix).

\section{Simon's problem}
Simon's algorithm~\cite{simon} is the first quantum algorithm that is demonstrated
to be exponentially faster than any probabilistic classical algorithm in solving a
black-box problem. In this problem~\cite{simon, hen}, there is an $n$-bit integer $a$ 
such that for any two $n$-bit inputs $i$, $j$, a black-box function $f:\{0,1\}^{n}\rightarrow
\{0,1\}^{n-1}$ outputs the integers $f\left( i\right) =f\left( j\right) $ if
and only if $i\oplus j=a$, here $\oplus $ denotes the bitwise XOR operation.
The task is to find $a$ by querying the function $f$. It requires $%
O(2^{n/2}) $ queries of $f$ for classical algorithms to solve this problem,
while Simon's algorithm solves the problem with $O(n)$ queries of $f$ on a
quantum computer.

The function in Simon's problem is a two-to-one mapping, two states $%
|i\rangle $, $|j\rangle $ with $i\oplus j=a$ form a set corresponds to an
integer in the set $\{0,1\}^{n-1}$. There are $2^{n-1}$ sets and each set
contains two CBS. The period $a$ can be calculated if we obtain the state of
a set. Let $h_{k}$ be the integer associated with the state $|k\rangle $~($%
k=0,\cdots ,2^{n}-1$), $h_{k}\in \left( 0,\cdots ,2^{n-1}-1\right) $, then $%
h_{i}=h_{j}$ if and only if $j=i\oplus a$. The Hamiltonian for Simon's
problem can be constructed by using the function $f$ as:
\begin{equation}
H_{\text{simon}}|k\rangle =h_{k}|k\rangle \text{.}
\end{equation}%
Our algorithm can be used to find the set associated with integer $0$. We
prepare a set of threshold values $\left\{
v_{1}=2^{n-2}-1,v_{2}=2^{n-3}-1,\ldots ,v_{n-1}=0\right\} $, and use $m=n-1$
oracles to construct Hamiltonians $H_{P_{i}}$ as in Eq.~($7$), the search
spaces of the Hamiltonians have the same nested structure as that of the
HSP. The search space of the problem is narrowed in a rate of $1/2$ in each
step by using the oracles based on the above threshold values. Then we
construct a sequence of Hamiltonians $H_{i}$ as shown in Eq.~($3$) and run
the algorithm to obtain the superposition state of two CBS in the set
associated with the integer $0$ in $n-1$ steps. The period $a$ is determined
in runtime $O(n)$.

\section{The graph isomorphism problem}
The GIP is considered as one of the few natural problems in the complexity class of NP
that could be classified as neither NP-complete nor P, for a review on the GIP, see 
Refs.~\cite{groh,babai}. The GIP is described as follows: given two undirected graphs 
$A_{1}(V_{1},E_{1})$ and $A_{2}(V_{2},E_{2})$ with vertex sets $V_{1}$, $V_{2}$, 
and edge sets $E_{1}$, $E_{2}$, respectively, and $\left\vert
V_{1}\right\vert =\left\vert V_{2}\right\vert =n$, the graph $A_{1}$ is
isomorphic to $A_{2}$, if there exists a bijection $\rho
:V_{1}\longrightarrow V_{2}$ such that for all $x,y\in V_{1}$, $(x,y)\in
E_{1}$ if and only if $(\rho x,\rho y)\in E_{2}$. The GIP can be reduced to
the graph automorphism problem, which is a relabeling of vertices of a graph
that preserves the graph edges. The graph automorphism problem for a graph $%
A(V,E)$ is to find a set of generators for the group of automorphisms $Aut(A)
$ of $A$.

The following reduction of the GIP to the graph automorphism problem is due
to Jozsa~\cite{Jozsa}. We construct a graph $A$ as disjoint union of graphs $%
A_{1}$ and $A_{2}$, having $2n$ vertices labeled $1,\ldots ,n$, $n+1,\ldots
,2n$ where $1,\ldots ,n$ label $A_{1}$ and $n+1,\ldots ,2n$ label $A_{2}$,
the automorphism group $K=Aut(A)$ embeds into the symmetric group $S_{2n}$.
Any automorphism of $A$ must either permute the vertices of $A_{1}$ and $%
A_{2}$ separately, or swap their vertices entirely. Let $Q$ denotes the
group $S_{n}\times S_{n}$, which contains all permutations that map $A_{1}$
and $A_{2}$ into themselves, separately, and let $\sigma =\left( 1\text{ }%
n+1\right) \left( 2\text{ }n+2\right) \ldots \left( n\text{ }2n\right) $ be
the permutation of $1,2,\ldots ,2n$ that swaps the vertices of $A_{1}$ and $%
A_{2}$ in their listed order. The automorphism group $K$ is a subgroup of
the group $Y=Q\cup \sigma Q$. If $A_{1}$ and $A_{2}$ are not isomorphic,
then $K=Aut(A)$ lies entirely in $Q$. Otherwise, if $A_{1}$ and $A_{2}$ are
isomorphic, then exactly half of the elements of $K$ are in $Q$, and the
other half are in $\sigma Q$. We can check if an element $\pi $ of $K$ lies
in $Q$ or $\sigma Q$ by evaluating $\pi (1)$. Therefore whether or not $%
A_{1} $ and $A_{2}$ are isomorphic can be determined with high probability
by randomly sampling the elements of $K$. For graph $A$, the automorphism
group $K$ is the hidden subgroup of the group $Y$. The function $%
f:Y\longrightarrow X$, where $X$ is the set of graphs created by
permutations of $A$ by $f(\pi )=\pi A$, is constant on the cosets of $K$ in $%
Y$, it separates the cosets of $K$ and is efficiently computable. The GIP is
reduced to a non-Abelian HSP where $f$ hides the automorphism group $K$ of
graph $A$, solving the HSP on the symmetric group leads to a solution to the
GIP~\cite{juj}.

Based on the above analysis, if the hidden subgroup $K$ is obtained, we can
check if an element $\pi $ of $K$ lies in $Q$ or $\sigma Q$ by evaluating $%
\pi (1)$ to determine if $A_{1}$ and $A_{2}$ are isomorphic with high
probability. Our algorithm can obtain the state $|K\rangle =\frac{1}{\sqrt{%
\left\vert K\right\vert }}\sum_{k_{i}\in K}|k_{i}\rangle $ of the hidden
subgroup $K$ of $Y$ efficiently. The group $Y$ has $\left\vert Y\right\vert
=2\left( n!\right) ^{2}$ group elements, they are mapped to at most $2\left(
n!\right) ^{2}$ integers from $1$ to $2\left( n!\right) ^{2}$ by applying
the function $f$. Using the oracle $O_{f}:|j\rangle |0\rangle
\longrightarrow |j\rangle |f\left( j\right) \rangle $, where the state $%
|j\rangle $ is associated with a permutation $\pi _{j}\in Y$, the
Hamiltonian of the GIP is constructed as%
\begin{equation}
H_{\text{GIP}}|j\rangle =h_{j}|j\rangle \text{,\ \ \ }j=1,\ldots ,2\left(
n!\right) ^{2}.
\end{equation}%
A set of threshold values $\left\{ v_{1}=\left( n!\right) ^{2},v_{2}=\frac{%
v_{1}}{2},\ldots ,v_{m}=\frac{v_{m-1}}{2}=1\right\} $ is used to construct $%
m=\log _{2}\left[ 2\left( n!\right) ^{2}\right] $ oracles. We construct the
Hamiltonians $H_{P_{i}}$ by using these oracles in the same way as in Eq.~($7
$), and the search spaces of the Hamiltonians have the nested structure, the
ratio $N_{i}/N_{i-1}$ for this problem is either one or ${1/2}$ in each
step. Then we construct the Hamiltonians $H_{i}$ as shown in Eq.~($3$) and
run the algorithm by setting the initial state of the problem as $|\psi
_{0}\rangle =\frac{1}{n!\sqrt{2}}\sum_{j=0}^{2\left( n!\right)
^{2}-1}|j\rangle $. The state $|K\rangle $ can be obtained in $m$ steps. By
performing measurement on the state $|K\rangle $, we can randomly obtain an
element of the subgroup ${K}$, and determine if $A_{1}$ and $A_{2}$ are
isomorphic with high probability by applying the element to the vertex $1$
of the graph $A_{1}$. The total runtime of the algorithm for solving the GIP
is proportional to $m$ and scales as $O(\log |S_{n}|)$.

\section{The poly($n$)-unique shortest vector problem}
The poly$(n)$-unique SVP is defined on a point lattice. An $n$-dimensional 
point lattice is a discrete subset of $\mathbb{R}^{n}$ closed under 
addition and substraction. It is generated by a set of integer linear combination 
of $n$ linearly independent vectors $\mathbf{b}_{1},\ldots ,\mathbf{b}_{n}$ as
\begin{equation}
B=\left\{ \sum_{i=1}^{n}x_{i}\mathbf{b}_{i}:x_{i}\in \mathbb{Z}\right\} ,
\end{equation}%
where $\mathbb{Z}$ denotes the integer set. The SVP is: given a lattice
described by the basis vectors, find the shortest nonzero vector in the
lattice. The $n$ dimensional SVP is extremely difficult, all known
algorithms for the SVP require exponential time in $n$. A vector is $f(n)$%
-unique if it is a factor of $f(n)$ shorter than all other nonparallel
vectors. Approximating the shortest vector to within a constant, that is,
the $O(1)$-unique SVP is known to be NP-hard under a randomized reduction~%
\cite{micc}. The $O(2^{n})$-unique SVP is solvable in polynomial time by
using the LLL algorithm~\cite{lens}. The decision version of the poly($n$%
)-unique SVP is known to be in NP and coNP~\cite{ahar,lyub}.

It has been found~\cite{regev} that the poly$(n)$-unique SVP can be reduced
to the two-point problem on lattice, and the two-point problem can be
reduced to the dihedral coset problem, which can be reduced to the dihedral
HSP. Therefore an efficient quantum algorithm for the HSP of the dihedral
group $D_{N}$ will provide an efficient way of solving the poly$(n)$-unique
SVP. The dihedral group $D_{N}$ has $2N$ elements, it is a group of
symmetries of an $N$-sided regular polygon. There are $2N$ ways one can
apply rotations or reflections in a distinct way. Every element of $D_{N}$
can be represented as a tuple $(r_{1},r_{2})$ where $r_{1}\in \left\{
0,1\right\} $ representing the number of reflections and $0\leq r_{2}<N$
representing the number of rotations. The dihedral coset problem is to find
a constant integer $l$, given a collection of states in form of%
\begin{equation}
\frac{1}{\sqrt{2}}\left( |0\rangle |x\rangle \!+\!|1\rangle |\left(
x\!+\!l\right) \text{ mod }N\rangle \right) \text{ \ }x\in \left\{ 0,\ldots
,N-1\right\} .
\end{equation}%
They can be thought of as cosets of the subgroup $K=\{(0,0),(1,l)\}$ in $%
D_{N}$. Solving the dihedral HSP by finding the subgroup $K$ results in a
solution to the dihedral coset problem, and finally a solution to the
poly($n$)-unique SVP.

In the dihedral HSP, a function $f$ is defined to hide the subgroup $K$ by
mapping $K$ to a constant~(say $0$), and each of the other $N-1$ cosets of $%
K $ to a distinct integer. The dihedral group $D_{N}$ has exponential
number~(in $\log N$) of cosets of order two, and is infeasible to solve by
classical algorithms. While the dihedral HSP has the same structure as that
of Simon's problem in our algorithm, i.e. each set contains two elements and
is mapped to a distinct integer. Therefore the state $|K\rangle =\frac{1}{%
\sqrt{2}}\left( |0\rangle |0\rangle \!+\!|1\rangle |\!l\rangle \right) $ of
the hidden subgroup $K$ can be obtained with runtime $O\left( \log N\right) $
by applying our algorithm in the same way as that of in solving Simon's
problem in the main text. The constant $l$ can be determined with high
probability by measuring the state $|K\rangle $, and solving the dihedral
coset problem, thus obtaining the solution to the poly$(n)$-unique SVP.

\section{Discussion}
We study the spectrum structure of the HSP, and find
that it can be reduced to the nested structured search problem, which is
solved efficiently through a multistep quantum computation process.
Therefore both the Abelian and non-Abelian HSP can be solved in the same way
as that of the nested structured search problem. Our algorithm shows that a
multistep quantum computation process using a number of oracles can be more
efficient than a quantum computation process of using one oracle. Some
problems that can be reduced to the HSP, e.g. the discrete logarithm problem
and the GIP, are believed to be in the complexity class of NP-intermediate,
and the HSP can be reduced to the nested structured search problem. It is
likely that the nested structured search problem is also in the complexity
class of NP-intermediate, we leave this as an open question for future study.

\begin{acknowledgements}
We thank H. Xiang, A. Miranowicz and F. Nori for helpful discussion.
This work was supported by the Fundamental Research Funds for the Central
Universities~(Grant No.~11913291000022), and the Natural Science Fundamental
Research Program of Shaanxi Province of China~(Grant No.~2022JM-021).
\end{acknowledgements}

\appendix

\begin{appendix}

\section{Application of the algorithm for the factoring problem}
Factoring an integer $Z=x\times y$ on a quantum computer can achieve
exponential speedup over the best known classical algorithm by using Shor's
algorithm~\cite{shor}. An integer $a<Z$ and co-prime with $Z$ is used in the
algorithm, the order of $a$ is defined as the smallest integer $r$ that
satisfies $a^{r}=1($mod $Z)$, which can be found efficiently by using the
order-finding algorithm through quantum Fourier transform. Then the factors
of $Z$ can be calculated as gcd$(a^{r/2}\pm 1,Z)$. The cost of Shor's
algorithm scales as $O(L^{3})$, where $L=\lceil \log _{2}^{Z}\rceil $.

Our algorithm~\cite{wyx} can also be applied for solving the factoring
problem. The Hamiltonian for factoring an integer $Z$ is defined as
\begin{equation}
H_{\text{FP}}|k\rangle =h_{k}|k\rangle =a^{k}(\text{mod }Z)|k\rangle ,\text{
}k=0,1,\ldots ,N-1,
\end{equation}%
where $N=2^{n}>Z$ contains a few periods of the integer $a$. The ground
state eigenvalue of $H_{\text{FP}}$ is $1$, and the corresponding
eigenstates are in form of $|p\cdot r\rangle $, where $p=0,1,\ldots ,\lfloor
\left( N-1\right) /r\rfloor $. The order $r$ of $a$ can be determined by
obtaining the ground state of the Hamiltonian $H_{\text{FP}}$. The
eigenvalues of $H_{\text{FP}}$ are integers that distribute uniformly in the
period of $a$, since $k$ ranges from $0$ to $N-1$ uniformly, and have the
same spectrum as shown in Fig.~$1$ of the main text. We construct a set of
threshold values $\left\{ {v_{1}=\lfloor \frac{Z}{2}\rfloor ,v_{2}=\lfloor
\frac{v_{1}}{2}\rfloor ,\ldots ,v_{L}=\lfloor \frac{v_{L-1}}{2}\rfloor =1}%
\right\} $ (it could be $v_{L-1}=\lfloor \frac{v_{L-2}}{2}\rfloor =1$, then
the set contains $L-1$ elements, for convenience, we assume there are $L$
elements). The Hamiltonians $H_{P_{i}}$ can be constructed by using $L$
oracles based on the threshold values as shown in Eq.~($7$) of the main
text. The search spaces of the Hamiltonians $H_{P_{i}}$ have the nested
structure and are reduced in a rate of about $1/2$. Then we construct a
sequence of intermediate Hamiltonians as shown in Eq.~($3$) of the main text
to form a Hamiltonian evolution path to the problem Hamiltonian $%
H_{P}=H_{P_{L}}$, and run the algorithm. Finally we obtain a superposition
state in form of $\frac{1}{\sqrt{N/r}}\sum_{p=0}^{N/r}|p\cdot r\rangle $.
The total runtime of the algorithm for obtaining the ground state of the
problem Hamiltonian $H_{P}$ is proportional to the number of steps $L$ of
the algorithm, and scales as $O(\log N)$. The order $r$ of $a$ can be
obtained by measuring the ground state of the problem Hamiltonian, thus
solving the factoring problem.

\section{Application of the algorithm for the discrete logarithm problem}
The discrete logarithm problem can be described as follows~\cite{childs}:
for a cyclic group $G$ generated by an element $g$, given an element $x\in G$,
find the discrete logarithm of $x$ with respect to $g$, $\log _{g}x$,
which is the smallest nonnegative integer $j$ such that $g^{j}=x$. This
problem can be transformed to a period-finding problem~\cite{nc}. Define a
function $f(x_{1},x_{2})$ $=a^{sx_{1}+x_{2}}($mod $N)$, where all the
variables are integers, $r$ is the smallest positive integer for which $%
a^{r}\left( \text{mod }N\right) =1$. This function is $2$-tuple periodic
where $f(x_{1}+l,x_{2}-ls)=f(x_{1},x_{2})$. Determining $s$ leads to a
solution to the discrete logarithm problem~\cite{nc}: given $a$ and $b=a^{s}$%
, find the integer $s$. By applying the quantum order-finding algorithm~\cite%
{dislog}, one can obtain the order $r$ of $a$ with one query of a quantum
black box $O_{f}$ that performs the unitary transform $O_{f}|x_{1}\rangle
|x_{2}\rangle |y\rangle \rightarrow |x_{1}\rangle |x_{2}\rangle |y\oplus
f\left( x_{1},x_{2}\right) \rangle $, and $O\left( \left\lceil \log
r\right\rceil ^{2}\right) $ other operations, then $s$ can be obtained from $%
r$.

Our algorithm can also be applied for solving the discrete logarithm
problem. For the periodic function $f\left( x_{1},x_{2}\right) $, we define
the Hamiltonian of the discrete logarithm problem as
\begin{eqnarray}
H_{\text{DLP}}|x_{1}\rangle |x_{2}\rangle \!
&=&\!h_{x_{1}x_{2}}|x_{1}\rangle |x_{2}\rangle   \notag \\
&=&b^{x_{1}}a^{x_{2}}(\text{mod }N)|x_{1}\rangle |x_{2}\rangle   \notag \\
&=&a^{sx_{1}+x_{2}}(\text{mod }N)|x_{1}\rangle |x_{2}\rangle ,
\end{eqnarray}%
where $x_{1},x_{2}=0,\ldots ,N-1$. The ground state eigenvalue of $H_{\text{%
DLP}}$ is $1$. The eigenvalues of $H_{\text{DLP}}$ distribute uniformly in
the period of $a$, since $x_{1}$ and $x_{2}$ range from $0$ to $N-1$
uniformly. We construct a set of threshold values $\left\{ {v_{1}=\lfloor
\frac{N}{2}\rfloor ,v_{2}=\lfloor \frac{v_{1}}{2}\rfloor ,\ldots
,v_{m}=\lfloor \frac{v_{m-1}}{2}\rfloor =1}\right\} $, where $m=\log _{2}N$.
The Hamiltonian $H_{P_{i}}$ can be constructed by using $m$ oracles in the
same way as in Eq.~($7$) of the main text. The search spaces of the
Hamiltonians $H_{P_{i}}$ have the nested structure and are reduced in a rate
of about $1/2$ in each step. Then we construct a Hamiltonian evolution path
as shown in Eq.~($3$) of the main text, and let $H_{P}=H_{P_{m}}$ and run
the algorithm. The ground state of $H_{\text{DLP}}$ in form of $\frac{1}{%
\sqrt{N/r}}\sum_{sx_{1}+x_{2}=kr}|x_{1}\rangle |x_{2}\rangle $ ($k=0,\ldots
,\lfloor \left( N-1\right) /r\rfloor $) can be obtained in $m$ steps. One
can calculate the integer $s$ by measuring the qubits to obtain states $%
|x_{1}\rangle $ and $|x_{2}\rangle $, thus solving the discrete logarithm
problem.

\section{Application of the algorithm for the period-finding problem and the
order-finding problem}
The period-finding problem can be described as follows~\cite{nc}: suppose $f$
is a periodic function producing a single bit as output such that $%
f(x+r)=f(x)$ for some unknown $0<r<2^{m}$, where $x,r\in \{0,1,\ldots \}$.
Given a quantum black box $O_f$ that performs the unitary transform $%
O_f|x\rangle |y\rangle \rightarrow |x\rangle |y\oplus f(x)\rangle $, how
many black box queries and other operations are required to determine $r$?
Quantum algorithm solves this problem using one query of the black box and $%
O(L^{2})$ other operations through the quantum Fourier transform~\cite{nc}.

By using the periodic function $f$, we can construct the Hamiltonian of the
periodic-finding problem as:%
\begin{equation}
H_{\text{PF}}|x\rangle =h_{x}|x\rangle \text{,\ \ \ }x=0,1,\ldots ,N
\end{equation}%
where $N=2^{m}$ is an integer that contains a few periods of $x$. We
construct a set of threshold values $\left\{ {v_{1}=\lfloor \frac{N}{2}%
\rfloor ,\ v_{2}=\lfloor \frac{v_{1}}{2}\rfloor ,\ldots ,v_{L}=\lfloor \frac{%
v_{m-1}}{2}\rfloor =1}\right\} $. The period-finding problem has the same
spectrum as shown in Fig. $1$ of the main text, the states $|x+kr\rangle $, $%
k=0,\ldots ,2^{m}/r$ have the same function value.

We can construct the Hamiltonian $H_{P_{i}}$ by using $m$ oracles in the
same way as in Eq.~($7$) of the main text, and a sequence of intermediate
Hamiltonians to reach the problem Hamiltonian $H_{P}=H_{P_{m}}$ as shown in
Eq.~($3$) of the main text, respectively. By running the algorithm in $m$
steps, we can obtain a superposition state {in form of $\frac{1}{\sqrt{N/r}}%
\sum_{k=0}^{N/r}|x+kr\rangle $} that corresponds to the ground state
eigenvalue of the function $f$. The period $r$ can be determined by
measuring the state.

The order-finding problem is as follows~\cite{nc}: for positive integers $y$
and $N$, $y<N$, that have no common factors, the order of $y$ modulo $N$ is
defined to be the least positive integer $r$, such that $y^{r}=1$(mod $N$).
The problem is to determine the order $r$ for some specified $y$ and $N$.
Order-finding is believed to be a hard problem on a classical computer, and
is used as the basis for the RSA crypto system. As shown in solving the
factoring problem, the order-finding problem can be solved efficiently by
using our algorithm in $L$ steps, with $L\equiv \log _{2}N$ as the number of
bits needed to specify $N$.
\end{appendix}

\end{document}